\documentclass[conference]{IEEEtran}

\makeatletter
\def\ps@IEEEtitlepagestyle{%
  \def\@oddfoot{\mycopyrightnotice}%
  \def\@oddhead{\hbox{}\@IEEEheaderstyle\leftmark\hfil}\relax
  \def\@evenhead{\@IEEEheaderstyle\hfil\leftmark\hbox{}}\relax
  \def\@evenfoot{}%
}
\def\mycopyrightnotice{%
  \begin{minipage}{\textwidth}
  \centering \scriptsize
\copyright 2021 IEEE.  Personal use of this material is permitted.  Permission from IEEE must be obtained for all other uses, in any current or future media, \\
including reprinting/republishing this material for advertising or promotional purposes, creating new collective works, for resale or redistribution to \\
servers or lists, or reuse of any copyrighted component of this work in other works. DOI 10.1109/SEED51797.2021.00012
  \end{minipage}
}
\makeatother

\newif\ifNotes\Notesfalse
\newif\ifAnon\Anonfalse
\newif\ifCamera\Camerafalse

\Notestrue 

\ifCamera
  \Notesfalse
  \Anonfalse
\fi

\IEEEoverridecommandlockouts
\usepackage{amsmath,amssymb,amsfonts}
\usepackage{algorithmic}
\usepackage{graphicx}
\usepackage{textcomp}
\usepackage[hyphens]{url}
\usepackage[numbers,sort]{natbib}
\usepackage{xcolor}
\usepackage{xspace}
\usepackage{hyperref}
\usepackage[nameinlink,capitalise]{cleveref}
\def\BibTeX{{\rm B\kern-.05em{\sc i\kern-.025em b}\kern-.08em
    T\kern-.1667em\lower.7ex\hbox{E}\kern-.125emX}}
\begin{document}

\title{Seeds of SEED: A Side-Channel Resilient Cache Skewed by a Linear Function over a Galois Field}

\author{\IEEEauthorblockN{Scott Constable}
\IEEEauthorblockA{\textit{Security and Privacy Research (SPR)}\\
\textit{Intel Corporation}\\
scott.d.constable@intel.com}
\and
\IEEEauthorblockN{Thomas Unterluggauer}
\IEEEauthorblockA{\textit{Security and Privacy Research (SPR)} \\
\textit{Intel Corporation}\\
thomas.unterluggauer@intel.com}
}
%\author{\IEEEauthorblockN{Anonymous}
%\IEEEauthorblockA{Anonymous Institution}
%}

\ifNotes
\newcommand{\colorcomment}[2]{\leavevmode\unskip\space{\color{#1}#2}\xspace}
\else
\newcommand{\colorcomment}[2]{\leavevmode\unskip\relax}
\fi

\renewcommand{\sectionautorefname}{Section}%
\renewcommand{\subsectionautorefname}{Section}%

\newcommand{\taggedcolorcomment}[3]{\colorcomment{#1}{[\textbf{#2}: #3]}}
\newcommand{\todo}[1]{\colorcomment{red}{[TODO: #1]}}
\newcommand{\scott}[1]{\taggedcolorcomment{red!50!black}{scott}{#1}}
\newcommand{\thomas}[1]{\taggedcolorcomment{blue}{thomas}{#1}}

\newcommand{\permutation}{$\Pi$\xspace}
\newcommand{\Zp}{\mathbb{Z}_p}
\newcommand{\Ztwon}{\mathbb{Z}_{2^n}}

\maketitle

\begin{abstract}
Consider a set-associative cache with $p^n$ sets and $p^n$ ways where $p$ is prime and $n>0$. Furthermore, assume that the cache may be shared among $p^n$ mutually distrusting principals that may use the \textsc{Prime+Probe} side-channel attack against one another; architecturally, these principals occupy separate security domains (for example, separate processes, virtual machines, sandboxes, etc.). This paper shows that there exists a linear skewing of cache sets over the Galois field $G_{p^n}$ that exhibits the following property: each cache set of each security domain intersects every cache set of every other security domain exactly once. Therefore, a random eviction from a single cache set in security domain $A$ may be observed via \textsc{Prime+Probe} in any of security domain $B$'s cache sets. This paper characterizes this linear skewing and describes how it can be implemented efficiently in hardware.
\end{abstract}


\section{Introduction}\label{sec:intro}

% FIXME: Lots of citations here

Side-channel attacks against set-associative caches have been studied for more than a decade. The recent disclosures of Spectre~\cite{KocherHFGGHHLM019}, Meltdown~\cite{Lipp0G0HFHMKGYH18}, and other transient execution attacks have only accelerated this research. Cache paritioning across ways~\cite{LiuGYMRHL16,KirianskyLADE18}, across sets~\cite{Kessler92} (known as \emph{coloring}), or individual lines~\cite{WangL07} can provide strong isolation and obviate side-channel leakage, but at the potential cost of reduced cache utilization~\cite{Dessouky20}. Other side-channel resilient cache proposals use randomization~\cite{LiuWML16} or cryptographic hashing~\cite{WernerUG0GM19} to obfuscate address-to-set and/or address-to-way mappings. These schemes may incur little or no performance penalty, yet they often fail to meet their intended security objectives~\cite{Bourgeat20,Purnal19}. To compensate for these shortcomings and provide convincing security, these caches must be periodically re-keyed, i.e., the set/way mapping must be redrawn to prevent the adversary from reverse engineering any given mapping and then extracting secrets~\cite{Qureshi18,WernerUG0GM19}. This causes overhead and limits the practicality of the cache.

This paper examines a radically different approach to designing side-channel resilient caches: instead of partitioning the cache or obfuscating its layout, let the entire cache be shared (no partitions) and allow the layout of the cache to be fixed at design time and publicly known; then, construct the layout to maximize the difficulty of a cache-based side-channel attack. There may be many possibilities for designing such a cache, though to the best of the authors' knowledge, this approach has not been considered in any prior work.

This paper introduces \textsc{GaloisCache}, a side-channel resilient cache that is skewed by a publicly known linear function that does not change. This particular linear function is parameterized by a security domain ID, and it minimizes the number of intersections between any given cache set in the victim's security domain and any given cache set in the adversary's security domain. Instead of integers, the sets and ways of \textsc{GaloisCache} are indexed by polynomials in a Galois field. Galois-field arithmetic is an area of immense importance in cryptography because addition and multiplication of Galois polynomials is usually efficient to implement in hardware. But in this work, Galois polynomials are primarily of interest because of one of their algebraic properties, that every Galois polynomial has a multiplicative inverse within the same Galois field. The fact that Galois arithmetic is efficient is merely a beneficial coincidence (see Section \ref{sec:impl}).

The many prior works that use randomization or hashing to obfuscate the cache layout are often accompanied by a security analysis based on probabilistic reasoning~\cite{Bourgeat20,Qureshi18,WernerUG0GM19,Qureshi19}. By contrast, the security of \textsc{GaloisCache} is predicated upon a single discrete mathematical property of the cache layout, and discrete mathematical properties are, in general, easier to scrutinize that probabilistic security models. This has also led the authors to quickly discover a novel kind of cache-based side-channel attack against \textsc{GaloisCache}, a kind of \emph{collusion attack} where multiple adversaries must collude to launch a side-channel attack against a victim that is sharing the same \textsc{GaloisCache}.

The contributions of this paper can be summarized as follows:
\begin{enumerate}
	\item \textsc{GaloisCache} is, to the best of the authors' knowledge, the first attempt to construct a side-channel resilient cache without using partitioning or set/way obfuscation techniques. The cache skewing function can also be embedded with low latency directly into the memory pipeline, using only a small set of XOR gates.
	\item The cache collusion attack is, to the best of the authors' knowledge, a novel kind of cache side-channel attack that may have implications for other cache designs that employ randomization, hashing, or skewing to obfuscate the cache layout.
\end{enumerate}

This work is organized as follows: \autoref{sec:background} gives an overview on the background, \autoref{sec:galoiscache} introduces \textsc{GaloisCache}, and \autoref{sec:conclusion} concludes this work.

\section{Background}
\label{sec:background}

Before describing \textsc{GaloisCache} in more detail, this section gives background on state-of-the-art caches, incidental channels, and previous work on side-channel resilient caches. 

\subsection{Caches}

Caches are essential parts of modern computing architectures as they bridge the performance gap between fast CPUs and slow DRAM. They store data fetched from memory closer to the CPU processing the data to allow for quicker responses when the same data is needed again. The smallest granularity that caches can store is determined by the size of an architecture's cache line, 64~bytes on current Intel systems. Caches may be organized in hierarchies, where smaller, faster first-level caches are placed closer to the CPU core and larger, slower last-level caches are placed closer to the memory. The majority of these caches is set-associative, which means that they are organized in sets and ways. Cache line addresses are statically mapped to a cache set through its address bits and can be placed in any of the respective cache set's ways. Caches use a replacement policy to determine a cache line to be evicted to memory (or the next-level cache in a cache hierarchy) when another cache line is inserted and no more cache lines are empty in the target cache set. Common replacement policies being used are variants of (Pseudo) Least Recently Used (LRU/PLRU)~\cite{Denning68} and random replacement. The latter picks a cache line in the target cache set to be evicted uniformly at random, whereas LRU/PLRU maintain state information per cache line or cache set to evict data that is unlikely to be used in the near future.

\subsection{Incidental Channels}

In a computing system, a \emph{channel} is a medium through which information (e.g., data) may be transmitted. When a channel is explicitly intended by the system designer to be used for the transfer of information, it is a \emph{legitimate channel}. All other channels are \emph{incidental channels}, i.e., their ability to transfer information is incidental to their intended purpose~\cite{intel21RefinedSpeculativeExecutionTerminology}. Examples of legitimate channels include shared memory, internal computer buses, and all forms of I/O. A simple example of an incidental channel is the CPU power state~\cite{uefi21ACPISpec}. If the CPU is in the ``Working" state, then its power draw will be relatively high; if the CPU is in the ``Sleeping" state, then its power draw will be relatively low.

Incidental channels can have implications for system security. If a malicious adversary has tapped the power supply to the CPU, then the adversary may be able to use power measurements to infer when a victim user is running an intensive task, such as playing a game or mining Bitcoin. In this case, the incidental channel is being used by the adversary as a \emph{side channel}: the victim user is unintentionally leaking information through the CPU's power state (the incidental channel's input) to the adversary who is monitoring the draw on the computer's power supply (the incidental channel's output). Alternatively, the adversary may control both input and output, in which case the incidental channel is being used as a \emph{covert channel}. For example, the adversary may use the ``Working" state to send a ``1"-bit over the channel, and the ``Sleeping" state to send a ``0"-bit over the channel. The adversary can use this technique to bypass a confinement policy~\cite{Lampson1973} intended to prevent sensitive information from escaping a computing system.

\subsection{Cache Incidental Channels}

The type of incidental channels concerned in this work are cache incidental channels. Cache incidental channels build upon the timing difference between requests hitting and missing in the cache. This intrinsic property of caches allows the adversary to infer information about the victim's cache activity. For example, a victim application on processor P0 may evict a cache line used by the adversary on another processor P1 back to memory. The adversary's next access to that memory region will then be measurably slower—a hint that the victim may have accessed memory corresponding to the evicted cache line. 
Eventually, this may allow an adversary such as a malicious OS, hypervisor, or user-space application, to capture sensitive information from a victim application through a side channel. The ability of software to prevent these \emph{side-channel vulnerabilities} is constrained by its limited view of microarchitectural platform state. In the previous example , the victim application cannot feasibly determine whether a cache line is present in a particular cache before it attempts to access data on that line. Set-associative caches have also been analyzed as high-bandwidth primitives to form covert channels~\cite{LiuYGHL15,Mengjia16,GrussMWM16}.

\begin{figure}
 \begin{center}\footnotesize
 \begin{tabular}{cc}
 \includegraphics[width=0.4\columnwidth]{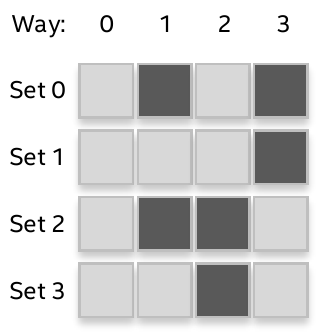} &  \includegraphics[width=0.4\columnwidth]{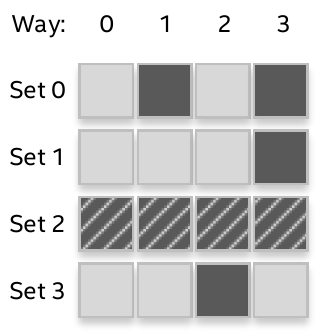} \\
 (a) Initial state & (b) \textsc{Prime} step \\
 \includegraphics[width=0.4\columnwidth]{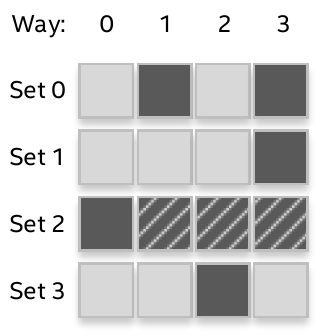} &  \includegraphics[width=0.4\columnwidth]{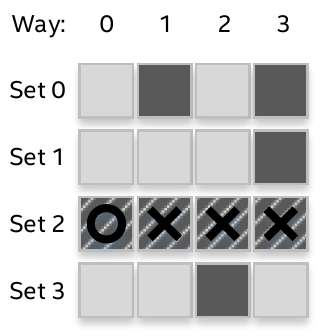} \\
 (c) Victim fill & (d) \textsc{Probe} step
 \end{tabular}
 \end{center}
\caption{\textsc{Prime+Probe} (solid dark: victim data; slashed dark: adversary data; hits in the ``probe'' step are shown as X's, misses shown as O's).}
\label{fig:primeprobe}
\end{figure}

There are multiple techniques to capture data leakage through cache side channels, such as \textsc{Prime+Probe}~\cite{OsvikST06} and \textsc{Flush+Reload}~\cite{YaromF14}. These exploit two main sources of leakage: (1) cache contention and (2) shared cache lines. \textsc{Prime+Probe}, for example, is a commonly cited technique to exploit cache contention. \autoref{fig:primeprobe} depicts the stages of a \textsc{Prime+Probe} \emph{cache-based side-channel attack} against a shared 4-way set-associative cache with 5 cache sets. The stages are as follows: (a) the victim runs for a while, loading some data into the cache; (b) the adversary “primes” a cache set by loading data into all of its lines; (c) the adversary waits for the victim to make more accesses; (d) the adversary “probes” the cache set by attempting to access the same data that was primed. If a primed line was evicted by the victim thread, then the adversary probe on that line will miss—and the probe will require more CPU cycles to complete. By repeating this attack against strategically chosen cache sets, the adversary can learn about the victim's memory access pattern and expose the victim's secrets. Namely, the static mapping of memory addresses to cache sets provides information about the addresses accessed by the victim, which attackers can use to, e.g., extract AES keys~\cite{OsvikST06}.

\subsection{Side-channel Resilient Caches}

There are three techniques for mitigating cache-based side-channel attacks that are generally regarded as robust~\cite{GrasRBG18}. They are (1) partitioning cache ways according to security domains~\cite{LiuGYMRHL16,KirianskyLADE18}; (2) partitioning cache sets according to security domains, a.k.a. “cache coloring”~\cite{ShiSCZ11,CostanLD16}; and (3) enforcing cache quiescence, i.e., that loads/evictions/flushes within a certain subset of the cache are either prevented or can be detected~\cite{GrussLSOHC17}.

\begin{figure}
 \begin{center}
 \includegraphics[width=\columnwidth]{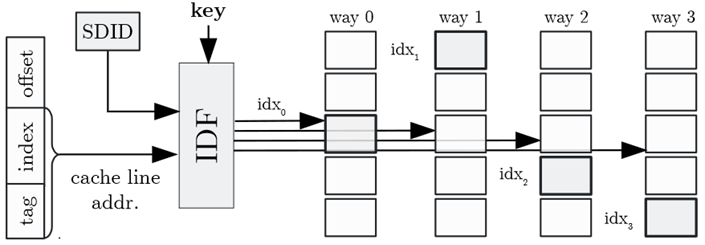}
\end{center}
\caption{A \textsc{ScatterCache} design for a 4-way set-associative cache~\cite{WernerUG0GM19}}
\label{fig:scattercache}
\end{figure} 

More recently, researchers proposed cache skewing~\cite{WernerUG0GM19,Qureshi19} as a technique to dynamically randomize cache sets in a set-associative cache to reduce side-channel leakage between security domains that share the cache. For instance, cache skewing may be applied to reduce leakage in the last-level cache (LLC) that is shared by all threads that execute on the same CPU. In \textsc{ScatterCache}~\cite{WernerUG0GM19}, each security domain is additionally distinguished by a Security Domain Identifier (SDID), which is used to derive the indices to access cache lines in each cache way. The researchers describe two basic approaches to implement this index derivation. The first is to use a keyed hash function (called Index Derivation Function or IDF) to assign a pseudo-random cache set (i.e., a set of indexes to be used to access each cache way) for each address based on a SDID and a key, as depicted in \autoref{fig:scattercache}. The second approach is to use SDID, key, and the address tag bits to differently permute the index bits used for each way. A particular algorithm for permuting the cache sets is not specified in the paper and must be designed according to the respective cache dimensions.

Although cache and way partitioning do provide strong separation between security domains, they substantially constrain the amount of memory that each security domain can access in the cache. For instance, if a cache is set/way partitioned equally among 8 security domains, then each security domain can only utilize $1/8$ of the cache. Moreover, when a security domain is data intensive, its data accesses will self-evict more frequently, degrading performance. When a security domain is not data intensive, then that domain’s cache partition will be underutilized, wasting valuable on-chip memory. Cache quiescence causes similar problems, as the lines that are being protected for a given security domain are unusable by other security domains.

Skewed cache designs such as \textsc{ScatterCache} largely solve the utilization problem, but they can also introduce other problems. For example, both \textsc{ScatterCache} variants require a pseudo-random hash/permutation function that may be computationally expensive, thus making its implementation in the timing sensitive memory pipeline challenging. The first variant of \textsc{ScatterCache} also invites the possibility of, for given tag bits, multiple set indices mapping to the same cache set, thus increasing the probability for self-evictions and degrading performance. In addition, the security of \textsc{ScatterCache} relies on the secrecy of the randomized mapping of addresses to cache lines. To ensure the secrecy of the mapping and thus side-channel resilience of \textsc{ScatterCache} over long periods of time, \textsc{ScatterCache} may occasionally need to be re-keyed, as otherwise an adversary may be able to recover details of the remapped cache sets~\cite{Qureshi19,PurnalGGV21}. This re-keying entails flushing or relocating lines in the entire cache, which is very expensive for larger cache structures such as the LLC.

\section{GaloisCache}\label{sec:galoiscache}

As outlined in \autoref{sec:background}, current approaches to protect caches against side-channel attacks based on partitioning are inflexible, whereas randomization-based designs may suffer from costly re-keying to keep the randomized mapping secret and guarantee long-term security. \textsc{GaloisCache} is a unique solution intended to address both problems by relying on a static and publicly known skewing function.

\subsection{Concept}

Define a \emph{security domain} to be a collection of threads, processes, VMs, etc. within which every thread/process/VM trusts every other thread/process/VM; any entity within a security domain implicitly distrusts every entity within every other security domain. The basic idea of this design is to permute the ways of a set-associative cache in a manner that establishes the following property: each cache set of each security domain intersects every cache set of every other security domain exactly once. This property, combined with random or pseudo-random replacement, renders cache-based side-channel attacks such as \textsc{Prime+Probe} substantially more difficult.

\subsection{Design}

\begin{figure}
 \begin{center}\footnotesize
 \begin{tabular}{cc}
 \includegraphics[width=0.45\columnwidth]{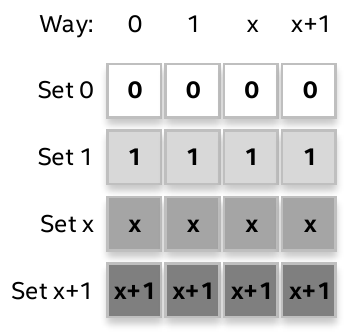} &  \includegraphics[width=0.45\columnwidth]{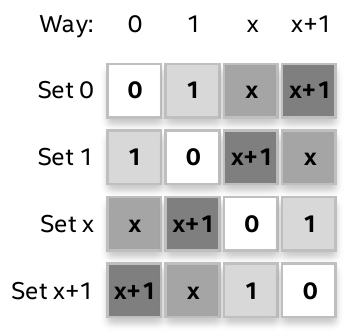} \\
 Security Domain $0$ & Security Domain $1$ \\
 \includegraphics[width=0.45\columnwidth]{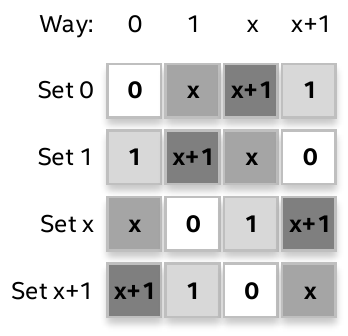} &  \includegraphics[width=0.45\columnwidth]{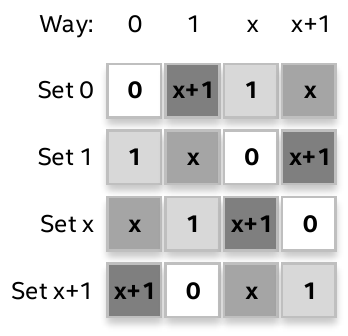} \\
 Security Domain $x$ & Security Domain $x+1$
 \end{tabular}
 \end{center}
\caption{Example security domain layout for \permutation\ over $GF(4)$. Cache sets are shaded to emphasize the skewed layout of each security domain.}
\label{fig:galois}
\end{figure}

The property that each cache set of each security domain intersects every cache set of every other security domain exactly once will hence be referred to as \emph{cache diagonalization}. Cache diagonalization can be realized by using a linear permutation as shown in the following in its general form: 
\begin{equation*}
 \Pi(t,s,w) = a \cdot s + b \cdot t \cdot w + c \;(\text{mod}\ R)
\end{equation*}
where $a$ is any non-zero constant, $s$ is the (input) cache set index, $b$ is any non-zero constant, $t$ is the security domain ID, $w$ is the way index, $c$ is any constant, and $R$ is the modulus over some Galois field $GF(p^n)$ where $p$ is prime and $n>0$. The output of \permutation is the permuted cache set index for set index $s$ at way $w$ for security domain $t$. \autoref{fig:galois} illustrates the output of \permutation for each of 4 security domains that share a cache with 4 sets and 4 ways (i.e., $p=2$, $n=2$). Each shade (corresponding to a unique Galois polynomial) represents a cache set that has been permuted by \permutation. Notice, for example, that each cache set in security domains 1, 2, and 3 intersects each cache set in security domain 0 exactly once.

\begin{figure}
 \begin{center}\footnotesize
 \begin{tabular}{cc}
 \includegraphics[width=0.45\columnwidth]{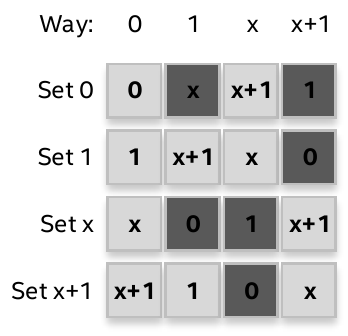} &  \includegraphics[width=0.45\columnwidth]{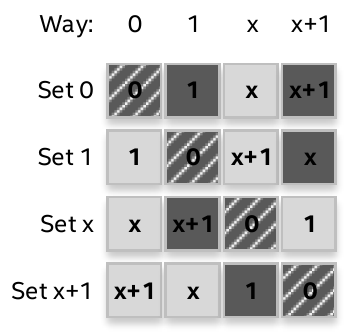} \\
 (a) Initial state (victim's layout) & (b) \textsc{Prime} (adversary's layout) \\
 \includegraphics[width=0.45\columnwidth]{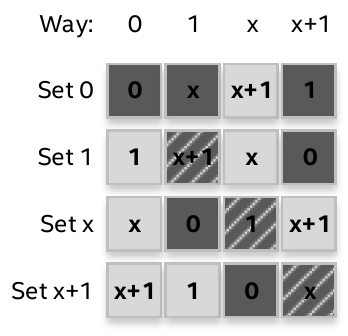} &  \includegraphics[width=0.45\columnwidth]{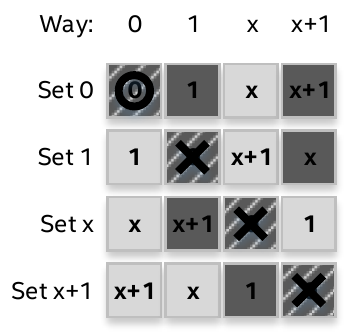} \\
 (c) Victim fill (victim's layout) & (d) \textsc{Probe} (adversary's layout)
 \end{tabular}
 \end{center}
\caption{Attempted \textsc{Prime+Probe}. The victim is security domain $x$ and the adversary is security domain $1$ (solid dark: victim data; slashed dark: adversary data; hits in the ``probe'' step shown as X's, misses shown as O's).}
\label{fig:galoispp}
\end{figure}

\autoref{fig:primeprobe} showed how a \textsc{Prime+Probe} attack can be carried out on a commodity set-associative cache. \autoref{fig:galoispp} shows how the same \textsc{Prime+Probe} attack would be mitigated by the cache diagonalization property of \textsc{GaloisCache}. Suppose that the adversary (security domain $1$) would like to detect whether the victim (security domain $1$) will load data whose address maps to set $0$ in the victim's security domain. The steps are as follows: (a) the victim  runs for a while, loading some data into the cache; (b) the adversary primes a cache set within its own security domain, e.g., cache set $0$; (c) the victim loads data that is not present in the cache and whose address maps to cache set $0$ within the victim's security domain, and way $0$ is randomly selected (with $\frac{1}{4}$ probability) for replacement; (d) the adversary probes cache set $0$ and observes one miss. However, the adversary can only deduce that \textit{some} cache line was filled by the victim\textemdash{}the adversary does not learn the corresponding set index within the victim domain. Consider: if the victim had filled data into set $1$, then with $\frac{1}{4}$ probability the adversary would also observe one miss in the probe step, in way $1$. If the victim had filled data into set $x$ then with $\frac{1}{4}$ probability the adversary would observe one miss in way $x$, and similarly a fill into set $x+1$ would be observed with $\frac{1}{4}$ probability as a miss in way $x+1$. Therefore, the standard \textsc{Prime+Probe} attack within a \textsc{GaloisCache} defined over $G_{p^n}$ will only tell the adversary whether the victim filled any line, with $1/{p^n}$ probability. Section \ref{sec:security} describes a more elaborate \textsc{Prime+Probe}-style attack that can leak address information, albeit with $1/{p^n}$ probability.

\subsection{Mathematical Foundation}

To build an intuitive understanding of how this design works, first consider the case where $n=1$, i.e., the Galois field $GF(p)$ is of prime order. In this case, the ring of cache set indexes $\Zp$ is isomorphic to $GF(p)$. One noteworthy consequence of this observation is that every non-zero element $e \in \Zp$ has a multiplicative inverse: $e^{-1} \in \Zp$.

Now consider arbitrary security domain IDs $t$ and $t'$ (i.e., $ t \neq t'$), and arbitrary cache set indexes $s,s' \in \Zp$. To establish the diagonalization property, it suffices to show that for fixed $a$, $b$, and $c$, the equation $\Pi(t,s,w)=\Pi(t',s',w)$ has a unique solution for $w$. That is, all cache sets in $t$ and $t'$ intersect along exactly one way. Hence solving for $w$:
\begin{align}
  \Pi(t,s,w) &= \Pi(t',s',w) \\
  a\cdot s + b \cdot t \cdot w + c	&=	a \cdot s' + b \cdot t' \cdot w + c \;(\text{mod}\ p) \label{eq:derivation_line2}\\
  b \cdot t \cdot w - b \cdot t' \cdot w	&=	a \cdot s' - a \cdot s \;(\text{mod}\ p) \label{eq:derivation_line3}\\
  b \cdot (t-t') \cdot w	&=	a \cdot (s'-s) \;(\text{mod}\ p) \label{eq:derivation_line4}\\
  w	&= a \cdot (s'-s) \cdot b^{-1} \cdot (t-t')^{-1}  \;(\text{mod}\ p) \label{eq:derivation_line5}
\end{align}
The justification for each of the steps is as follows: \ref{eq:derivation_line2} by unfolding \permutation; \ref{eq:derivation_line3} and \ref{eq:derivation_line4} by basic modular arithmetic; and \ref{eq:derivation_line5} since b is non-zero it has a multiplicative inverse $b^{-1} \in \Zp$, and similarly for $(t-t')$ because $t \neq t'$.

One other useful property of \permutation\ is that it is a bijection of the cache set indexes within each way. This is important because it establishes that for each security domain all of its cache sets do not overlap one another—thus ensuring that each security domain can utilize the entire cache, and that the frequency of self-evictions is minimal.

Next, consider the complementary case where $n \neq 1$, and in particular with $p=2$. Although the ring of cache set indexes $\Ztwon$ is not isomorphic to $GF(2^n)$, there is a bijection between the two structures: for each $i \in \Ztwon$, interpret $i$'s sequence of binary digits as the coefficients of a polynomial of degree $n-1$. For example, cache set index $42=\text{0b00101010}$ corresponds to the polynomial $x^5+x^3+x$ in $GF(2^8)$. Additions in this field correspond to XORs of the polynomial coefficients (because the characteristic of the field is 2), and multiplications are ordinary polynomial multiplications, modulo a reducing polynomial. These operations over finite fields can be found in other cryptographic areas of computer science including encryption (e.g., AES, GCM) and error correction (e.g., Reed-Solomon), with fast algorithms for computing them.

The permutation function \permutation\ is similarly applicable to all other finite fields, and the argument given above for its validity over $GF(p)$ is identical to the argument for the more general case of $GF(p^n)$. The only distinction is that all of the variables and constants are interpreted as polynomials, and the modulus $R$ must be suitable for $GF(p^n)$ (such a modulus is guaranteed to exist). Finally, note that although it is theoretically possible to implement \permutation over $GF(p^n)$ for $p>2$ and $n>1$, the arithmetic operations become much more complicated to implement in hardware. The most likely implementations of cache diagonalization would have $p=2$, i.e., the number of sets and ways is simply a power of 2.

In general, a caching structure that uses \permutation\ to permute its $p^n$ cache sets must also have $p^n$ ways, and has the capacity to allow up to $p^n$ distinct security domains to share the caching structure without violating the cache diagonalization property.

\subsection{Implementation}\label{sec:impl}

The most relevant implementation parameter is the latency of the permutation $\Pi(t,s,w)$ as this permutation needs to be computed on every cache request and be embedded into the memory pipeline. To allow for efficient, low-latency implementations of the permutation $\Pi(t,s,w)$, the cache diagonalization can be instantiated with the parameters $p=2$ and $n>1$, i.e., the number of sets and ways is a power of 2. In this case, additions are done using bitwise XOR and hence feature very low hardware complexity. Multiplications, on the other hand, are composed of both AND and XOR gates, but for a carefully chosen irreducible polynomial (= modulus $R$) feature a low critical path as well, which has been shown before for implementations of, e.g., AES MixColumns~\cite{feldhofer2005}.

To outline the implementation efficiency of $\Pi(t,s,w)$, consider the scenario where $a=1$ and $b=1$. In this case, an implementation of \textsc{GaloisCache} only needs to perform the finite-field multiplications $t \cdot w \; (\text{mod}\ R)$ for all $w$. At a high level, these finite-field multiplications consists of two parts: a multiplication $t \cdot w$  and a reduction mod $R$. While generic finite-field multiplications in $GF(2^n)$ require AND gates to do the multiplication part $t \cdot w$, the cache architecture, its parameters and the set of values for $w$ are fixed at design time of the circuit. As a result, a circuit performing the multiplication $t \cdot w$ does not need any AND gates but can compute its result by simply XORing shifted instances of the input $t$, with a maximum latency of $n-1$ XOR gates. For the reduction part, depending on the output of the multiplication, shifted instances of the irreducible polynomial $R$ need to be XORed to the multiplication output. However, for wisely chosen irreducible polynomials, the latency is only slightly increased compared to plain polynomial multiplication. \autoref{tab:galoisfieldpolynomials} gives an overview on such irreducible polynomials and the respective cache dimensions. 

\begin{table}
	\caption{Irreducible polynomials of Galois fields and possible cache configurations.}
	\label{tab:galoisfieldpolynomials}
	\begin{center}
		\begin{tabular} { c c c }
			\textbf{Galois Field} & \textbf{Irreducible Polynomial} & \begin{tabular}[b]{@{}c@{}}\textbf{Cache Dimensions} \\ \textbf{(sets x ways)}\end{tabular} \\
      $GF(2^3)$ & $x^3+x^1+1$ & $8 \times 8$ \\
      $GF(2^4)$ & $x^4+x^1+1$ & $16 \times 16$ \\
      $GF(2^5)$ & $x^5+x^2+1$ & $32 \times 32$ \\
      $GF(2^6)$ & $x^6+x^1+1$ & $64 \times 64$ \\
      $GF(2^7)$ & $x^7+x^1+1$ & $128 \times 128$	
		\end{tabular}
	\end{center}
\end{table}

Note that for $b \neq 1$, the hardware can easily avoid the multiplication with $b$ in the critical path by precomputing $b \cdot t$ for each security domain, keeping the result in dedicated hardware registers and using these precomputed results in the following memory requests. For a constant $a \neq 1$, the multiplication $a \cdot s \; (\text{mod}\ R)$  can be efficiently implemented in the same manner as $t \cdot w \; (\text{mod}\ R)$. Because $a$ is fixed at design time, the multiplication $a \cdot s \; (\text{mod}\ R)$ as well features a short critical path consisting of a small number of XOR gates. Note that this multiplication $a \cdot s \; (\text{mod}\ R)$ does not increase the overall critical path of the permutation $\Pi(t,s,w)$ as the multiplication can be done in parallel to the multiplications $t \cdot w \; (\text{mod}\ R)$ for all $w$. Overall, the properties of the permutation $\Pi(t,s,w)$ allow it to be implemented with low latency and hence its circuit to be embedded directly into the memory pipeline. 

The most restrictive aspect of \textsc{GaloisCache} is its requirement that the cache itself must be square, i.e., that the number of sets must equal the number of ways. This restriction can be relaxed by ``stacking" $2^n$ \textsc{GaloisCache}s. If a single adversary is able to precisely monitor contention within each of the $2^n$ caches, then the adversary may be able to infer up to $n$ bits from a single victim access among the caches.

\subsection{Security Analysis and Cache Collusion Attacks}
\label{sec:security}
\label{sec:collusion}

Assume that the \textsc{GaloisCache} implementation uses random or pseudo-random replacement, and recall the cache diagonalization property: each cache set of each security domain intersects every cache set of every other security domain exactly once. Now consider the effect of a single eviction triggered by an operation within security domain $t$. From the perspective of $t$, the line was evicted from a random way within a single cache set. Yet from the perspective of any other security domain $t'$, by diagonalization a line may have been evicted from a random set. Therefore $t'$ cannot infer any information from the eviction, other than that some cache line was evicted by some other security domain. Most importantly, $t'$ cannot learn any of the address bits corresponding to the evicted line; this is the basis for classical cache-based side-channel attacks such as \textsc{Prime+Probe}~\cite{OsvikST06}, and also for transient execution attacks like Spectre~\cite{KocherHFGGHHLM019}. However, \textsc{GaloisCache} is susceptible to a new type of attack where \emph{multiple} malicious security domains may collude to infer more information about a victim access. 

\begin{figure}
 \begin{center}\footnotesize
 \begin{tabular}{cc}
 \includegraphics[width=0.45\columnwidth]{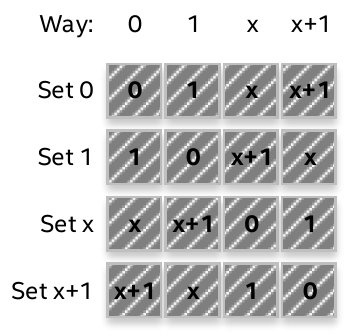} &  \includegraphics[width=0.45\columnwidth]{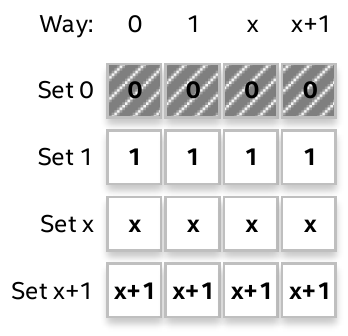} \\
(a) Security Domain $1$ & (b) Security Domain $0$ \\
 \includegraphics[width=0.45\columnwidth]{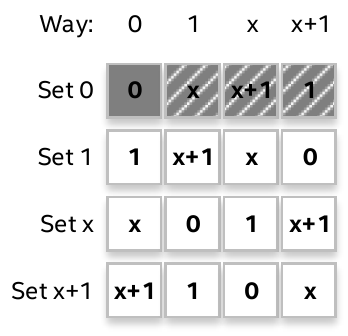} &  \includegraphics[width=0.45\columnwidth]{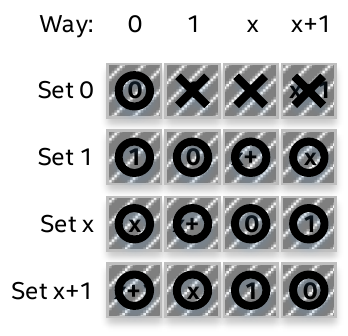} \\
 (c) Security Domain $x$ & (d) Security Domain $1$
 \end{tabular}
 \end{center}
\caption{Security Domains $1$ and $0$ collude against Security Domain $x$ (hits in the ``probe" step are shown as X's, misses are shown as O's)}
\label{fig:collusion}
\end{figure}

As mentioned in Section \ref{sec:intro}, a victim security domain in a \textsc{GaloisCache} may be vulnerable to a novel kind of \textsc{Prime+Probe} variant if at least two other security domains are able to collude and synchronize their cache accesses with each other and those of the victim. \autoref{fig:collusion} illustrates the steps required for colluding adversarial security domains $0$ and $1$ to detect with $1/p^n$ probability a single cache access made by the victim security domain $x$. The example revisits the $GF(4)$ \textsc{GaloisCache} illustrated above in \autoref{fig:galois}. Each of the four steps in the figure is shown from the perspective of the security domain that is accessing the cache in that step. White squares show data belonging to security domain $0$, dark slashed squares show data belonging to security domain $1$, and dark solid squares show data belonging to security domain $x$.

The steps to carry out the attack are as follows: (a) security domain $1$ primes the entire cache by filling all of its own cache sets; (b) security domain $0$ fills three of its own cache sets to evict exactly three lines from each set that was primed by security domain $1$, thus security domain $1$ is left with a single line in each of its own cache sets; (c) the victim security domain $x$ accesses memory that maps to its cache set $0$, and with probability $\frac{1}{4}$ security domain $1$'s remaining cache line in its cache set $0$ is evicted; (d) security domain $1$ probes the entire cache. If security domain $1$ records $4$ misses in its cache set $0$, then the victim must have accessed data mapping to its cache set $0$. Otherwise, either the victim did not access data in that set, or it did and the eviction occurred in another way with probability $\frac{3}{4}$.

Though the collusion attack demonstrates that \textsc{GaloisCache} is not impervious to side-channel attacks, \textsc{GaloisCache} does raise the bar for \textsc{Prime+Probe}-style techniques, especially compared to commodity set-associative caches. Consider: the collusion attack (a) requires the first adversary to prime and probe the entire cache, (b) requires at least one of the adversaries to be able to influence scheduling, and (c) is highly susceptible to noise from other processes since the attack cannot be localized to a small region of the cache. It is open to future work to explore whether and how the concept of collusion attacks could be improved by extending to a larger number of colluding domains.

\section{Conclusion}
\label{sec:conclusion}

This paper introduced \textsc{GaloisCache}, a novel side-channel resilient cache design that does not rely on partitioning or an obfuscated layout to mitigate side-channel attacks such as \textsc{Prime+Probe}. \textsc{GaloisCache} is also susceptible to a novel kind of collusion side-channel attack wherein two or more adversaries must coordinate a cache eviction strategy to infer a victim's cache access pattern. Although \textsc{GaloisCache} is vulnerable to this new kind of attack, the authors hope that this innovative design philosophy for side-channel resilient caches could inspire new research in this area.

% Unlikely to be approved by legal
%Side-channel attacks including \textsc{Prime+Probe}~{OsvikST06} and \textsc{Flush+Reload}~\cite{YaromF14} have been a concern of Intel and its customers for over a decade, but to date Intel has not provided an adequate solution. More recently, Spectre-style attacks [10] have used CPU caches to construct \emph{covert channels} that transmit program secrets from the victim to the adversary. This invention comprehensively defeats both categories of attacks, and thus will allow ISVs and end users to have more trust in the security of Intel’s CPUs.

\bibliographystyle{IEEEtranSN}
\bibliography{galois_cache}

\end{document}